\documentclass[a4paper,11pt]{article}
\pdfoutput=1 % if your are submitting a pdflatex (i.e. if you have
\usepackage{jcappub} 
\usepackage[T1]{fontenc} 

\usepackage{natbib}
\usepackage{cancel}
\usepackage{pifont} 
\usepackage{amsthm}         
\usepackage{amsmath} 	   
\usepackage{amssymb}       
\usepackage{amsfonts}
\usepackage{graphicx} 	    
\usepackage{subcaption} 	    
\usepackage{verbatim}
\usepackage{epsfig,bbm}
\usepackage{color}
\usepackage{colortbl}
\usepackage{float}
\usepackage{multirow}
\usepackage{lipsum}

\definecolor{babyblue}{rgb}{0.54, 0.81, 0.94}
\definecolor{corn}{rgb}{0.98, 0.93, 0.36}

\allowdisplaybreaks

\title{The robustness of slow contraction and the shape of the scalar field potential}

\author[a,b]{Timo Kist,} 
\affiliation[a]{Max Planck Institute for Gravitational Physics, 30167 Hanover, Germany}
\affiliation[b]{Fakultät für Physik, Georg-August-Universität Göttingen, 37077 Göttingen, Germany}
\author[c,1]{Anna Ijjas}
\note{Corresponding Author}

\affiliation[c]{Center for Cosmology and Particle Physics,  New York University, New York, NY 10003, USA}

\emailAdd{ijjas@nyu.edu}

\abstract{
We use numerical relativity simulations to explore the conditions for a canonical scalar field $\phi$ minimally coupled to Einstein gravity to generate an extended phase of slow contraction that robustly smooths the universe for a wide range of initial conditions and then sets the conditions for a graceful exit stage. We show that to achieve robustness it suffices that the potential $V(\phi)$ is negative and $M_{\rm Pl}|V_{,\phi}/V|\gtrsim5$ during the smoothing phase. We also show that, to exit slow contraction, the potential must have a  minimum. Beyond the minimum, we find no constraint on the uphill slope including the possibility of  ending on a positive potential plateau or a local minimum with $V_{\rm min}>0$. 
Our study establishes ultralocality for a wide range of potentials  as a key both to robust smoothing and to graceful exit.
}


\begin{document}
\maketitle 
\raggedbottom

\section{Introduction}
\label{sec_intro}

The striking simplicity of our large-scale universe remains a major puzzle for theoretical cosmology \cite{Turner:2018bcg}. 
Slow contraction -- a primordial phase that connects to the hot expansion stage through a cosmological bounce -- has been proposed as a dynamical mechanism to homogenize, isotropize and flatten (henceforth: smooth) the universe on super-Hubble scales \cite{Erickson:2003zm}. A key characteristic of slow contraction is the relationship between physical distances that evolve as the Friedmann-Robertson-Walker  (FRW) scale factor $a$, and those that evolve as the Hubble radius $|H^{-1}|$, which is the scale of causal connectedness. As the universe contracts, $a$ and $|H^{-1}|$ shrink at different rates, {\it i.e.},
\begin{equation}
\label{def-slow-contraction}
a \propto \left|H^{-1}\right|^{1/\epsilon},
\end{equation}
where $\epsilon=(3/2)(1+p/\rho)$ is the equation of state of the homogeneous energy component that dominates the total energy density and has pressure $p$ and energy density $\rho$.
For slow contraction where $\epsilon \gg 3$, the distance between two objects decreases much more slowly than the rate at which the Hubble radius shrinks.  An important consequence is that features originating on sub-Hubble scales, {\it e.g.}, primordial quantum fluctuations with initial wavelengths much smaller than the Hubble radius, end up being extended over scales exponentially larger than the Hubble radius by the end of the smoothing process, as required to explain the observed cosmic microwave background (CMB) fluctuations.

In a series of recent studies,  using the tools of mathematical and numerical relativity, it has been demonstrated that slow contraction  is a robust, rapid, and universal smoother (for a review, see \cite{Ijjas:2022qsv}). 
That is, slow contraction is achieved for a wide range of initial conditions (robustness) including those that lie far outside the perturbative regime of flat FRW spacetimes such 
as highly inhomogeneous, anisotropic and curved initial geometries.
Smoothness is typically achieved within ${\cal O}(1)$ $e$-folds of contraction of the Hubble radius (rapidity). 
In addition and perhaps more surprisingly, the numerical exploration has revealed that smoothing through slow contraction is universal \cite{Ijjas:2021gkf}: each spacetime point both within and outside any Hubble volume evolves independently towards the flat FRW state. This happens because, in contracting relativistic spacetimes, gradients quickly become negligible and do not contribute to the evolution (ultralocality).  The smooth FRW state is reached by each spacetime point because it is the only stable fixed point of the ultralocal system that has a large basin of attraction \cite{Ijjas:2021wml}. Notably, smoothing is not restricted to spacetime points that originate from the same Hubble volume. This is radically different from smoothing through inflation \cite{Ijjas:2022qsv}. 

Thus far, all numerical relativity studies of slow contraction involved a canonical scalar field $\phi$ that is minimally-coupled to Einstein gravity and has a negative exponential potential, 
\begin{equation}
\label{exponential-potential}
V(\phi) = -V_0 e^{-\phi/m},
\end{equation}
where $V_0$ and $m$ are positive constants and for these studies the potential was unbounded below. The scalar field acts as a microphysical source for a macroscopic equation of state, 
\begin{equation}
\epsilon \equiv 3\times \frac{\frac12 \phi'^2}{\frac12 \phi'^2 + V(\phi)} \gg3,
\label{eps-def}
\end{equation}
that triggers slow contraction as defined in Eq.~\eqref{def-slow-contraction}, where prime denotes differentiation w.r.t. the FRW time coordinate $\tau$.

In this paper, we lay the groundwork for model building. 
We extend earlier studies \cite{Cook:2020oaj,Ijjas:2020dws,Ijjas:2021gkf,Ijjas:2021wml} and elaborate the general conditions for a potential to achieve robust smoothing through slow contraction. Furthermore, we examine whether bounding the potential from below affects the robustness of smoothing. Potentials with global minima are of interest when constructing realistic scenarios since they connect straightforwardly to  an end of the smoothing slow contraction phase. 

The paper is organized as follows. We start with briefly outlining the numerical scheme and defining the procedure that we use to set the initial conditions in Sec.~\ref{sec:num}. We then demonstrate in Sec.~\ref{sec:downhill} that, to achieve robust smoothing,
it suffices that  the scalar field potential $V(\phi)$ is negative  and $M_{\rm Pl}|V_{,\phi}/V|\gtrsim5$ during the smoothing phase. (Here and throughout, $M_{\rm Pl}\equiv1/\sqrt{8\pi G_{\rm N}}$ denotes the reduced Planck mass with $G_{\rm N}$ being Newton's constant.) In Sec.~\ref{sec:uphill}, we show that, for the smoothing phase to end, the potential must be bounded from below and have a global minimum. Notably, we show that, due to the combination of robust and rapid smoothing and the rapidly growing Hubble anti-friction, completion of the smoothing phase is independent of the uphill slope.  Reaching the potential minimum, the scalar field quickly climbs uphill because the large Hubble anti-friction keeps increasing the field's kinetic energy density, which can enable the field to reach a positive potential energy plateau (or local minimum) that can terminate the slow contraction phase and connect smoothly to the hot expansion phase. In Sec.~\ref{sec:concl}, we conclude by pointing to possible future directions of research.

\section{Numerical scheme and initial conditions}
\label{sec:num}

To study the effect of various scalar field potential shapes on the robustness and rapidity of slow contraction, we numerically solve the full 3+1 dimensional Einstein-scalar field equations,
\begin{eqnarray}
\label{E-eq1}
R_{\mu\nu} - \frac12 g_{\mu\nu}R &=& \nabla_{\mu}\phi\nabla_{\nu}\phi -  g_{\mu\nu}\left( \frac12\nabla_{\lambda}\phi\nabla^{\lambda}\phi +V(\phi) \right),\\
\label{E-eq2}
\Box \phi &=& V_{,\phi},
\end{eqnarray}
where $R_{\mu\nu}$ is the Ricci tensor, $R=R_{\mu}{}^{\mu}$ is the Ricci scalar and $\Box=\nabla^{\lambda}\nabla_{\lambda}$ denotes the covariant d'Alembertian. 
%Throughout, we express microphysical quantities in reduced Planck units, $8\pi G_{\rm N} = 1$, with $G_{\rm N}$ being Newton's constant.
We use the same numerical scheme we employed for previous non-perturbative analyses of slow contraction, as fully detailed in Ref.~\cite{Ijjas:2021gkf}. Hence, in the following, we only walk through the essential aspects to keep the presentation self-contained. 
As described below, our scheme is particularly well-suited to the situation of slow contraction since it enables us to track the evolution for several hundreds of $e$-folds of contraction of the Hubble radius without encountering instabilities or singular behavior for a  wide range of initial conditions including those that lie outside the perturbative regime of FRW spacetimes. 

\subsection{Numerical scheme}
\label{sec:scheme}

An appropriate numerical general relativity scheme involves  a particular formulation of the field equations combined with a special gauge choice that leads to a system 
of coupled, non-linear partial differential equations (PDEs) that can then be solved using standard finite-difference or spectral methods; see, {\it e.g.}, \cite{Baumgarte:2021skc}. 
For example, the orthonormal tetrad form of the Einstein-scalar equations~(\ref{E-eq1}-\ref{E-eq2}) that underlies our numerical scheme  represents each space-time point through a set of four 4-vectors (tetrads); for the complete set of evolution and constraint equations, see Appendix~\ref{app:scheme}. The local 4-metric is flat (Minkowski) everywhere such that the geometric variables, which we evolve by way of the field equations, reduce to the sixteen tetrad vector components $E_a{}^i$ and the eighteen Ricci rotation coefficients $\gamma_{abc}$. The $\gamma_{abc}$ define how the tetrad is deformed when moving from one point to another. 

We specify the tetrad frame gauge by fixing six of the Ricci rotation coefficients such that no non-physical rotations are introduced (Fermi-propagation) and the time-like tetrad is normal to space-like hypersurfaces (hypersurface-orthogonality), {\it i.e.}, our tetrad gauge is chosen such that it defines both a particular frame and a particular foliation of spacetime into space-like hypersurfaces.
As a consequence of our frame gauge choice, the  remaining twelve Ricci rotation coefficients, which yield a complete set of geometric variables, represent physical quantities: the six components of the symmetric extrinsic 3-curvature tensor $K_{ab}$ and the six components of the symmetric part of the intrinsic (or spatial) 3-curvature tensor $N_{ab}$. 

To turn the tetrad equations into PDEs that we then numerically solve,  it is necessary to re-express directional derivatives along tetrads as partial derivatives along coordinate directions. To this end, we specify a particular coordinate gauge. Again, to avoid introducing unphysical gauge effects and obtain the formulation that readily relates to observables, we fix the shift vector such that the spatial coordinates are co-moving both with the tetrad frame and the associated foliation.  
Finally, we fix the lapse by requiring that spatial hypersurfaces of constant time are constant mean curvature (CMC) hypersurfaces. Here, the mean curvature $\Theta^{-1}$ is given by the trace of the extrinsic curvature, $\Theta^{-1}\equiv \frac13 K_a{}^a$. Note that in the homogeneous limit, the inverse mean curvature $\Theta$ coincides with the Hubble radius $|H^{-1}|$.

Our coordinate gauge choice is especially advantageous to simulate contracting spacetimes for two reasons. First, by our CMC gauge choice, the mean curvature is homogeneous and monotonic. We can therefore rescale the coordinate time $t$ to track the inverse mean curvature $\Theta$,
\begin{equation}
\label{time-def}
e^t = \frac13 \Theta,
\end{equation}
such that $N_{\Theta}\equiv t_0-t$ measures the number of $e$-folds of contraction in $\Theta$ between $t_0$ and $t$.
Note that, for slow contraction to explain the observed large-scale universe as well as the primordial fluctuation spectrum as seen in the CMB, the flat FRW state must be reached by $t-t_{\rm end}=60$, where $t_{\rm end}$ marks the end of slow contraction.
This is because the perturbation modes observed in the CMB must be generated on a flat FRW background over the course of the last 60 $e$-folds of slow contraction \cite{Cook:2020oaj}.

In addition, reaching the putative curvature (or big bang) singularity takes infinite coordinate time, {\it i.e.}, with $t$ running from zero to negative infinity, $\Theta\to0$ as $t\to-\infty$, but for all finite values of $t$ dynamical variables remain finite. That means, we can achieve stable evolution for arbitrary many $e$-folds of contraction independent of the actual bounce mechanism that eventually connects to the hot expanding phase.

Second, we can rescale all dynamical variables by normalizing them with the mean curvature $\Theta$,
\begin{eqnarray}
\label{cal-N-def}
N &\rightarrow& {\cal N}\equiv N \times \Theta^{-1},\\
\{E_a{}^i, \Sigma_{ab}, n_{ab}, A_b,  S_a\} &\rightarrow&  \{\bar{E}_a{}^i, \bar{\Sigma}_{ab} , \bar{n}_{ab}, \bar{A}_b, \bar{S}_a  \}
\equiv \{E_a{}^i, \Sigma_{ab}, n_{ab}, A_b, S_a\} / \Theta^{-1}  \,,
\\
\label{Vbar-def}
V &\rightarrow& \bar{V} \equiv V / \Theta^{-2}, 
\end{eqnarray}
where $\Sigma_{ab}$ is the trace-free part of the extrinsic curvature $K_{ab}$, $n_{ab}$ and $A_b$ are the  symmetric and anti-symmetric part of the intrinsic curvature $N_{ab}$, respectively,  $S_a$ is the scalar field gradient and bar denotes normalization by the mean curvature $\Theta^{-1}$ on constant time hypersurfaces (henceforth, Hubble-normalization).  Using dimensionless Hubble-normalized variables, we avoid numerical stiffness issues that we would expect given the exponentially different rates at which physical scales and the mean curvature evolve and that would limit the number of $e$-folds over which the code can run.

Furthermore, there exists a special set of Hubble-normalized variables, a combination of the geometric and scalar field variables, 
\begin{align}
\label{omega-s}
\Omega_s &= \;\tfrac{1}{6}\bar{\Sigma}^{ab}\bar{\Sigma}_{ab},  
\\
\label{omega-k}
\Omega_k &= -\tfrac{2}{3}{\bar{E}{}_a}^i\partial_i\bar{A}^a +\bar{A}^a\bar{A}_a +\tfrac{1}{6}\bar{n}^{ab}\bar{n}_{ab} -\tfrac{1}{12}({\bar{n}{}^c}_c)^2,
\\
\label{omega-phi}
\Omega_{\phi} &= \;\tfrac{1}{6}\bar{W}^2 + \tfrac{1}{6}\bar{S}^a\bar{S}_a + \tfrac{1}{3} \bar{V},
\end{align}
which we use to track whether and how smoothing is achieved. Here, $\bar{W}\equiv {\cal N}^{-1}\partial_t\phi$ denotes the Hubble-normalized scalar field velocity.
The $\Omega_i$ represent the fractional contribution of component $i$ (anisotropy, spatial curvature, scalar field) to the total energy density. Complete smoothness, {\it i.e.}, the flat FRW state, is reached when $\Omega_{\phi}=1$ and $\Omega_s, \Omega_k =0$.
Note that, in the homogeneous limit, the $\Omega_i$ coincide with the dimensionless Friedmann variables commonly used in the cosmology literature.

\subsection{Initial conditions}
\label{sec:init_cond}

The numerical scheme must be supplemented by initial conditions that satisfy a set of constraint equations~(\ref{constraintG}-\ref{constraintCOM}), {\it i.e.}, the projection of the Einstein-scalar field equations on a spatial hypersurface at some initial time $t_0$.
Since the field equations propagate these constraints, constraint satisfying initial data ensure energy and momentum conservation throughout the evolution.

Our scheme enables us to study a wide range of initial conditions including those that lie outside the perturbative regime of flat FRW spacetimes because, when specifying the initial data, we utilize all the degrees of freedom left after ensuring constant satisfaction.

More precisely, as common in numerical relativity studies, we adapt the York method \cite{York:1971hw} and define the spatial metric of the initial $t_0$-hypersurface to be conformally-flat,  
\begin{align}
    g_{ij}(t_0,\vec{x}) = \psi^4(t_0,\vec{x})\;\delta_{ij}, 
\label{eq:conf-flat}
\end{align}
where $\psi(t_0,\vec{x})$ the conformal factor.  
Together with the mean curvature $\Theta_0^{-1}$ at $t_0$ that we freely specify,  $g_{ij}(t_0)$ fixes the components of the spatial curvature tensor,
\begin{align}
    \bar{n}_{ab}(t_0,\vec{x}) &= 0, 
 \\
    \bar{A}_b(t_0,\vec{x}) &= -2\psi^{-1}(t_0,\vec{x})\;{\bar{E}{}_b}^i(t_0,\vec{x})\;\partial_i\psi(t_0,\vec{x}),
\label{eq:a0}
\end{align}
as well as the tetrad vector components,
\begin{align}
    {\bar{E}{}_a}^i(t_0,\vec{x}) = \psi^{-2}(t_0,\vec{x})\;\Theta_0^{-1}\;{\delta_a}^i.
\end{align}
Note that picking a conformally-flat initial metric is not a true restriction on the initial data. Rather, it is a practical device to ensure the constraints are satisfied. However, unlike constraint satisfaction, which is propagated by the Einstein-scalar equations, conformal flatness is broken within only a few integration steps, as verified by our numerous simulations.

At the same time, the York method enables us to  semi-analytically define the initial scalar field distribution,
\begin{align}
\phi(t_0,\vec{x}) &= f_x\cos(n_x x+h_x) + f_y\cos(n_y x+h_y) + \phi_0,
\label{eq:phi0} 
\end{align}
its conformally-rescaled velocity, 
\begin{align}
Q(t_0,\vec{x}) &= \Theta_0\left(q_x\cos(m_x x+d_x) + q_y\cos(m_y y+d_y) + Q_0\right),
\label{eq:q0}
\end{align}
as well as the divergence-free part of the conformally-rescaled anisotropy (or shear) tensor, 
\begin{align}
    Z_{a b}^{0} (t_0,\vec{x}) =
\begin{pmatrix}
b_{2}+c_{2} \cos y && \xi && \kappa_{1}+c_{1} \cos y \\
\xi && b_{1}+a_{1} \cos x && \kappa_{2}+a_{2} \cos x \\
\kappa_{1}+c_{1} \cos y &{\;}&  \kappa_{2}+a_{2} \cos x &{\;}& -b_{1}-b_{2}-a_{1} \cos x-c_{2} \cos y
    \end{pmatrix}.
\label{eq:z_ab0}
\end{align}
Here, $Q\equiv \psi^6\bar{W}$ and $Z_{ab}\equiv \psi^6\bar{\Sigma}_{ab}$ are the conformally rescaled scalar field velocity and shear tensor, respectively, specified through the parameters $\phi_0, Q_0, f_x, f_y, q_x, q_y, n_x, n_y, m_x, m_y$, $h_x$, $h_y$, $d_x, d_y$, $a_1, a_2, b_1, b_2, c_1, c_2, \kappa_1, \kappa_2$ and $\xi$ which we freely and independently set for each simulation run. 
The sinusoidal form of the spatial variations reflects that the boundary conditions are periodic, $0\leq x, y \leq 2\pi$, with $0$ and $2\pi$ identified. Although we only show a single mode for the shear and two modes for the field's velocity, they can be replaced by a sum of different fourier modes with different amplitudes, wavenumbers and phases.

The initial data is completed by numerically computing the conformal factor, $\psi(t_0,\vec{x})$, and the rest of the shear tensor, $Z_{ab} - Z^0_{ab}$, using the Hamiltonian and momentum constraints, respectively.

\subsection{Numerical implementation}
\label{sec:num_1d_exp}

The numerical analysis presented in this paper extends the results of Refs.~\cite{Cook:2020oaj,Ijjas:2020dws,Ijjas:2021gkf,Ijjas:2021wml}. To evolve the hyperbolic-elliptic system of partial differential equations~(\ref{eq-E-ai-Hn}) - (\ref{Neqn}), we discretize the equations using second order accurate spatial derivatives and a three-step method for time integration employing the Iterated Crank-Nicolson algorithm.  At each sub-step, we first solve the elliptic equation~\eqref{Neqn} through a multigrid V-cycle method and then update the hyperbolic equations (\ref{eq-E-ai-Hn}) - (\ref{eq-barS-Hn}) to the next Iterated Crank-Nicolson sub-step. In the simulations presented below, we use a grid of 1024 points with $\Delta x = 2\pi/1024$ and a Courant factor of 0.5. 

To demonstrate the convergence of our code, the error and convergence was analyzed for a broad range of examples using the same methods as detailed in the Appendices of Refs.~\cite{Ijjas:2020dws,Ijjas:2021gkf}. Our code shows no
signs of numerical instability and exhibits clear second order convergence at early times. At later times when a smooth, ultralocal spacetime develops, we empirically see the convergence improve to third order.

In order to compare whether and how the robustness and rapidity of smoothing is affected by the shape of the scalar field potential, we consider the same type of initial conditions as employed in Refs.~\cite{Ijjas:2020dws} and \cite{Ijjas:2021gkf}. In general, as specified below, we consider initial conditions that lie in the non-perturbative regime of flat FRW spacetimes. For example, 
we adjust the amplitude $q_x$ of the initial scalar field velocity fluctuations proportionally to the slope of the scalar field potential $|\bar{V}_{,\phi}/\bar{V}|$ such that $q_x\sim10\;\%$ of the flat FRW attractor value $Q_\mathrm{attr} \sim |\bar{V}_{,\phi}/\bar{V}|$. 
In addition, we choose the vacuum part of the initial shear tensor given by Eq.\eqref{eq:z_ab0} such that the initial shear density parameter $\Omega_s$ is the dominant contribution to the total energy density.

\section{Robustness and the downhill slope} 
\label{sec:downhill}

Non-perturbative, numerical relativity robustness studies of slow contraction \cite{Cook:2020oaj,Ijjas:2020dws,Ijjas:2021gkf,Ijjas:2021wml} so far have all involved a pure negative exponential potential that is unbounded below. One of the goals of this paper is to systematically explore the conditions on the scalar field potential to achieve robust and rapid smoothing. We do this by first presenting three representative cases involving a steep negative power-law potential, a negative super-exponential potential, and a shallow negative exponential potential. Then, using the type of dynamical systems analysis that we developed in Ref.~\cite{Ijjas:2020dws}, we show that, to achieve robust smoothing,
it suffices that  the scalar field potential $V(\phi)$ is negative and $M_{\rm Pl}|V_{,\phi}/V|\gtrsim5$ during the smoothing stage. 

\subsection{Worked examples}
\label{examples-smoothing}

To study whether and how the potential shape affects the robustness and rapidity of smoothing during slow contraction, we performed a large set of simulations involving a wide range of initial conditions and scalar field potentials with different signs, amplitudes and slopes. In the following, we will summarize our findings using representative examples. 

%FigureSmoothingOmegas
\begin{figure}%
    \centering
\includegraphics[width=.95\linewidth]{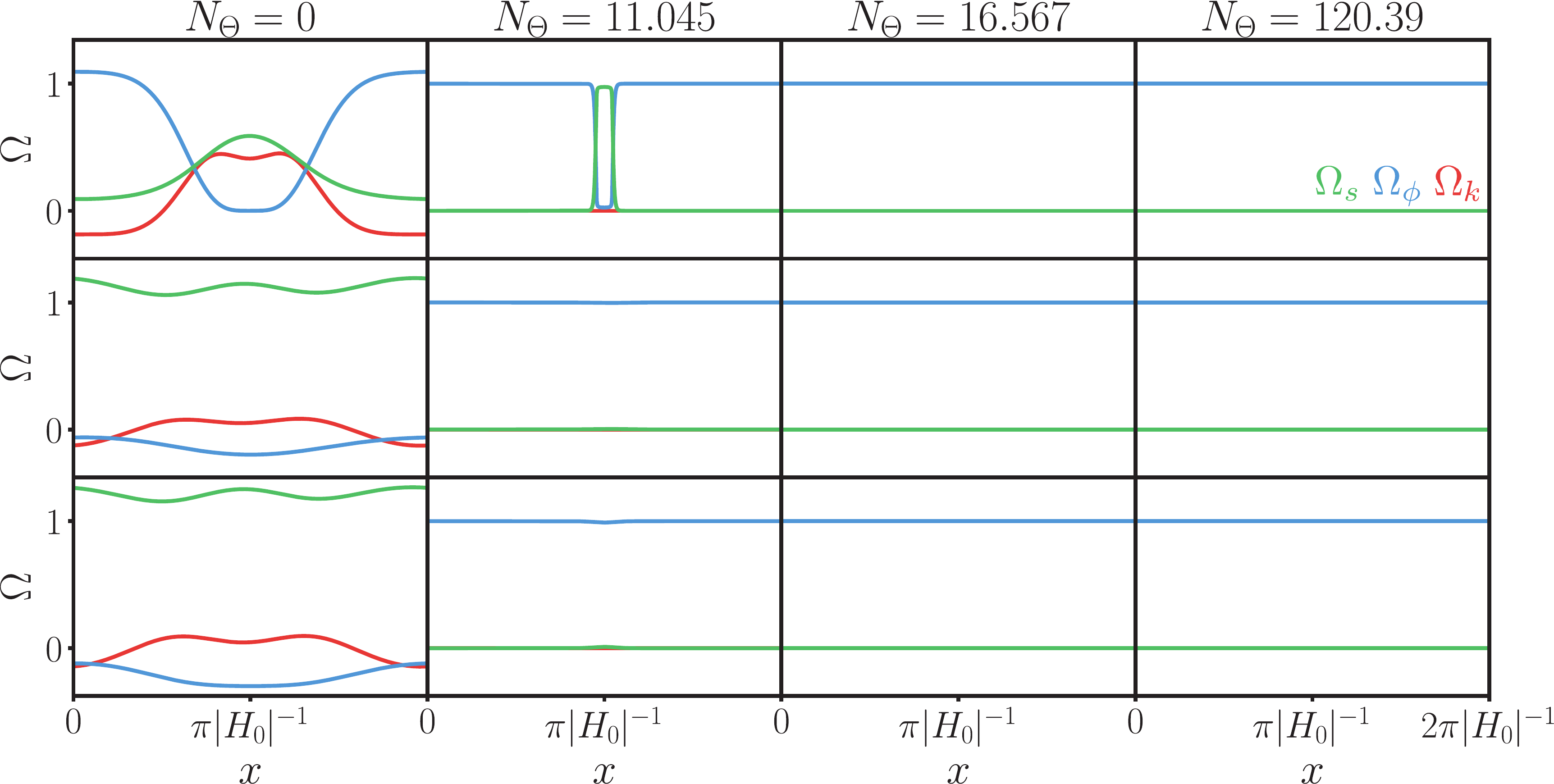}%
    \caption{Snapshots of the Hubble-normalized energy density in shear $\Omega_s$ (green), curvature $\Omega_k$ (red), and scalar field $\Omega_{\phi}$ (blue) as a function of the number of $e$-folds of contraction $N_{\Theta}$ in the inverse mean curvature $\Theta$ corresponding to a negative power-law potential (upper row), a negative super-exponential potential (middle row), and a steep negative exponential potential with $m=0.2$ (lower row). The initial states are highly curved, inhomogeneous and anisotropic and only differ in the initial scalar field velocity, as specified in Sec.~\ref{examples-smoothing}. Each of the three potentials leads to complete smoothing within less than 20 $e$-folds of contraction.}%
    \label{smoothing-omegas}%
\end{figure}
\begin{figure}[tb!]%
\centering
\includegraphics[width=\linewidth]{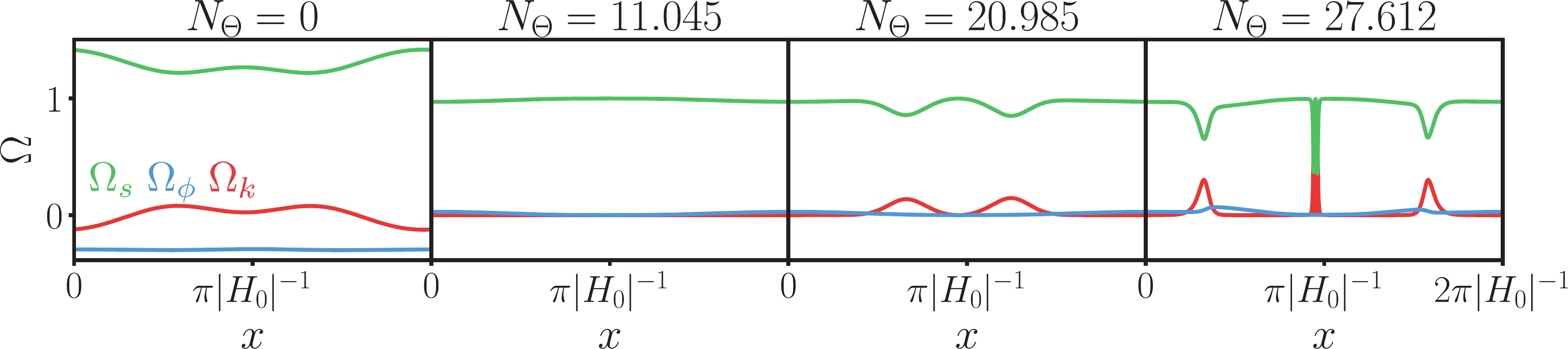}%
\caption{Snapshots of the Hubble-normalized energy density in shear $\Omega_s$ (green), curvature $\Omega_k$ (red), and scalar field $\Omega_{\phi}$ (blue) as a function of the number of $e$-folds of contraction in the inverse mean curvature $\Theta$ corresponding to a shallow negative exponential potential with $m=0.5$. The potential is not sufficiently steep to source a smoothing slow contraction phase but instead leads to a (degenerate) Kasner-like final state in which spacetime is continuously being bumped from one Kasner-like state to another.}%
\label{nosmoothing-omegas}%
\end{figure}

Figure~\ref{smoothing-omegas} shows snapshots of the evolution of the shear, spatial curvature and scalar field density parameters $\Omega_s, \Omega_k$ and $\Omega_{\phi}$ defined in Eqs.~(\ref{omega-s}-\ref{omega-phi}) for three different potential shapes: $V(\phi)=-V_0 \phi^n$ with $n=50$ (upper row), $V(\phi)=-V_0 \exp(\phi^2/M^2)$ with $M^2=1/3$ (middle row) and, as a reference case, $V(\phi)=-V_0 \exp(-\phi/m)$ with $m=0.2$ (lower row). In Figure~\ref{nosmoothing-omegas}, we present the evolution of the density parameters for a shallow negative exponential $V(\phi)=-V_0 \exp(-\phi/m)$ with $m=0.5$. For each potential, we set the coefficient to $V_0=0.1$ (in units of the mean curvature $\Theta$ at $t_0$).   
The three potentials corresponding to Fig.~\ref{smoothing-omegas} lead to complete smoothing ($\Omega_{\phi}=1$, $\Omega_s$,$ \Omega_k=0$) by $N_{\Theta}\simeq17$ (power-law), $N_{\Theta}\simeq11$ (super-exponential), and, $N_{\Theta}\simeq12$ (exponential), respectively, where $N_{\Theta}$ is equal to the number of $e$-folds of contraction of $\Theta$. The shallow negative exponential potential corresponding to Fig.~\ref{nosmoothing-omegas} does not lead to complete smoothing. Instead, some parts of the original Hubble-volume (the entire simulation box) are smoothed and some end in an anisotropic Kasner-like state ($\Omega_s=$ const. $\neq0$), locally developing chaotic mixmaster behavior \cite{Erickson:2003zm,Garfinkle:2008ei}. The case in Fig.~\ref{nosmoothing-omegas} is also representative of positive potentials and the case of a free scalar with no potential.
%FigureSmoothingOmegas

To evaluate their effect on the robustness of smoothing, we independently varied all parameters corresponding to the initial conditions in Eqs.~(\ref{eq:phi0}-\ref{eq:z_ab0}) from zero to ${\cal O}(1)$. Here and henceforth, we keep those parameters fixed that do not have any significant effect on robustness.
That is, in each of the four cases, the divergence-free part of the Hubble-normalized shear tensor $\bar{Z}_{ab}^0$ defined in Eq.~\eqref{eq:z_ab0} is set to 
\begin{align}
    Z_{a b}^{0} (t_0,\vec{x}) =
\begin{pmatrix}
b_{2} && \xi && 0 \\
\xi && b_{1}+a_{1} \cos (x+\phi_x) && a_{2} \cos (x+\phi_x) \\
0 &{\;}& a_{2} \cos (x+\phi_x) &{\;}& -b_{1}-b_{2}-a_{1} \cos (x+\phi_x)
    \end{pmatrix},
\label{eq:z_ab0-1D}
\end{align}
where $a_1=0.5$, $a_2=0.5$, $b_1=-0.15$, $b_2=1.8$, $\xi=0.01$ and $\phi_x=0.15$; the scalar field distribution is set to $\phi(t_0, {\bf x})=0$; and the period and shift of  the sinusoidal spatial variations of the scalar field velocity are fixed by setting $m_x=1, d_x=0$  and $m_y, d_y=0$. 

Note that, for the simulations presented in this Section, all deviations from homogeneity are along a single spatial direction. As shown in Ref.~\cite{Ijjas:2021wml} and below in Sec.~\ref{sec:(2+1)D}, this kind of initial condition is the least favorable for slow contraction to start. Intriguingly, inhomogeneities along two or three spatial directions are more favorable because they mean less symmetric initial conditions and more ways to carry the system towards the flat FRW attractor fixed point and away from the Kasner-like fixed point that appears to have a small basin of attraction.

We found that the key parameters that affect the robustness and the rapidity of smoothing are the sign and steepness of the scalar field potential, the field's average initial velocity $Q_0$ (in units of the mean curvature $\Theta$ at $t_0$) and the magnitude of the velocity fluctuation mode $q_x$ corresponding to the wavenumber $m_x$ as defined in Eq.~\eqref{eq:q0}. Our finding extends and generalizes the results of Refs.~\cite{Cook:2020oaj,Ijjas:2020dws}, which only considered pure exponential potentials. 
For the four cases that we present here, $Q_0$ and $q_x$ are specified for each potential separately as follows: 
For the super-exponential potential, $Q_0=1, q_x=0.3$; for the exponential potential with $m=0.2$, $Q_0=0.6, q_x=0.5$; for the power-law potential, $Q_0=5, q_x=5$; and   for the exponential potential with $m=0.5$, $Q_0=0, q_x=0.2$. Our convention is that $Q_0>0$ corresponds to a scalar field rolling downhill (towards more negative values of the potential). For the three potentials corresponding to Fig.~\ref{smoothing-omegas},  $q_x$ is adjusted to the slope such that the initial velocity fluctuation is non-perturbatively far from the flat FRW attractor value,  $Q_{\rm attr} \sim |\bar{V}_{,\phi}/\bar{V}|$; for example, $q_x \sim 0.1Q_{\rm attr}$. For the power-law and exponential potentials, the average initial velocity $Q_0$ is chosen such that it corresponds to the smallest value for which complete smoothing is reached. For the super-exponential, $Q_0$ is the smallest value for which complete smoothing is reached more rapidly than in the case of the exponential potential.

%
%FigureSN^-1
\begin{figure}%
    \centering
\includegraphics[width=0.95\linewidth]{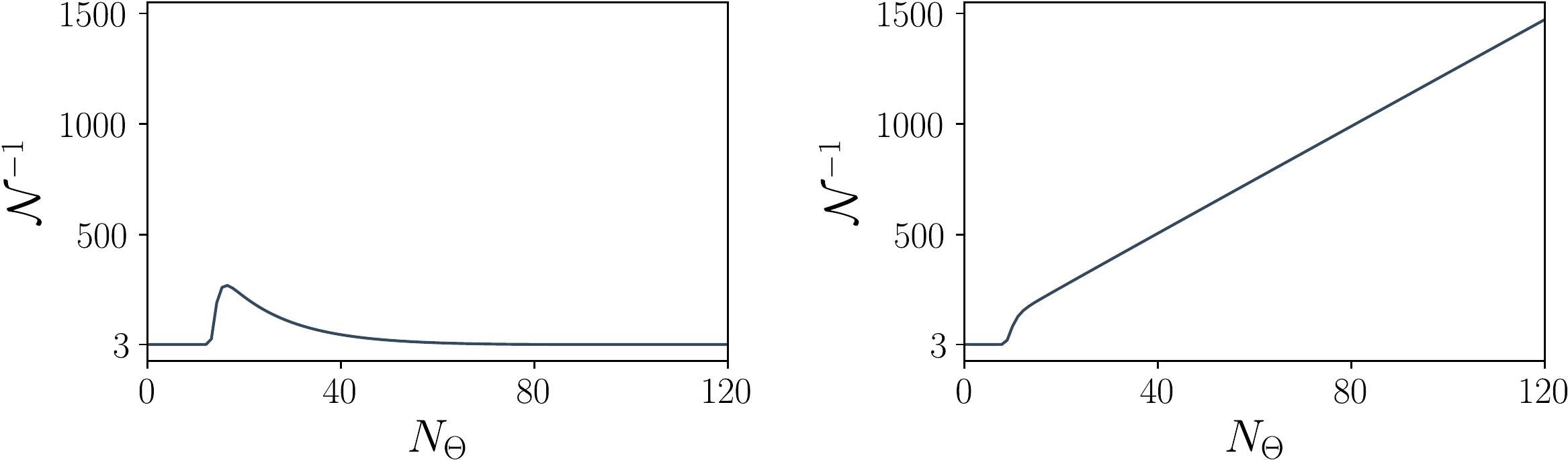}%
    \caption{Evolution of the Hubble-normalized lapse ${\cal N}^{-1}$ as a function of the number of $e$-folds of contraction $N_{\Theta}$ in the inverse mean curvature $\Theta$ corresponding to a negative power-law potential (left panel) and a negative super-exponential potential (right panel) for a select spatial point $x=\pi$ that starts out in a highly curved, inhomogeneous and anisotropic state. Once the smoothing phase is reached, ${\cal N}^{-1}$ coincides with the effective equation of state associated with the scalar field $\phi$ defined in Eq.~\eqref{eps-def}.}%
    \label{fig:sn}%
\end{figure}
Although robust smoothing could be achieved with power-law and super-exponential potentials as well, the overall dynamics is rather different from what we observe in the case of the pure exponential potential (middle row of Fig.~\ref{smoothing-omegas}). The difference can be illustrated, {\it e.g.}, through the evolution of the Hubble-normalized lapse.  
Figure~\ref{fig:sn} depicts the inverse Hubble-normalized lapse ${\cal N}^{-1}$ as a function of $t$ corresponding to the case of power-law and super-exponential potentials in Fig.~\ref{smoothing-omegas} for a representative spatial point $x=\pi$ that starts out in a highly curved, inhomogeneous and anisotropic region. In the case of a pure negative exponential, ${\cal N}^{-1}=1/(2\,m^2)$ during the smoothing phase. In the case of a power-law potential, ${\cal N}^{-1}$ first steeply approaches a large value ($\sim 250$) but then, having reached a maximum ($N_{\Theta}\simeq20$), ${\cal N}^{-1}$ relatively quickly (by $N_{\Theta}\simeq40$) falls back to 3 at which point the smoothing phase comes to an end. Yet, due to the rapidity of smoothing, we do not observe any deviations from homogeneity, isotropy or flatness at any later point, even when running the simulations for over 100 or more $e$-folds. In the case of the super-exponential, ${\cal N}^{-1}$ grows linearly as slow contraction proceeds. In particular, the smoothing case can last arbitrarily long just like the smoothing phase sourced by exponential potentials.

\subsection{Potential shape from dynamical systems analysis}

Having demonstrated numerically that sufficiently steep negative potentials all lead to an extended period of smoothing slow contraction that is robust to a wide range of initial conditions, including those that lie outside the perturbative regime of FRW spacetimes, we now show analytically that, to achieve robust smoothing,
it suffices that  the scalar field potential $V(\phi)$ is negative and $M_{\rm Pl}|V_{,\phi}/V|\gtrsim5$ during the smoothing stage. To this end we adapt the methods developed in the first systematic robustness study of slow contraction~\cite{Ijjas:2020dws}.

Ref.~\cite{Ijjas:2020dws} provides a dynamical systems method to identify the attractor fixed points of the coupled Einstein-scalar equations in the ultralocal limit when the macroscopic equation of state is sourced by a negative exponential potential. In our numerous simulations that involved negative as well as positive potentials ranging from power-law to exponential to super-exponential, we have confirmed that in general the system rapidly converges to an ultralocal ($\bar{E}_a{}^i, \bar{A}_{b},\bar{S}_a\simeq0$) state that is spatially-curved ($\bar{n}_{ab}\neq0$) and anisotropic ($\bar{\Sigma}_{ab}\neq0$). This means, the attractor fixed points of the coupled, non-linear Einstein-scalar PDE system are the same as the attractor fixed points of the coupled, linear ODE system (see Appendix~\ref{app:ODE}) that describes the evolution in the ultralocal limit.

Since, in the ultralocal limit, the momentum constraint~\eqref{constraintC}, which reduces to a simple algebraic relation: 
\begin{equation}
\epsilon _a{}^{b c} \bar{n}_b{}^d \bar{\Sigma}_{cd}=0, 
\end{equation}
implies that the shear and spatial curvature tensors commute,
it is straightforward to verify (see, Appendix~B in Ref.~\cite{Ijjas:2020dws}) that the essential time evolution of the Einstein-scalar ODE system is encapsulated by the {\it eigenvalue ODE system},
\begin{eqnarray}
%sigma1
\label{eq-sigma1-ul}
\dot{\sigma}_1 &=&   \Big(1 - 3{\cal N} \Big) \sigma _1
- {\textstyle \frac13 }  \, {\cal N}\left(   \big(2\nu_1 - \nu_2 - \nu_3 \big) \nu_1 - \big(  \nu_2- \nu_3  \big)^2 \right)
,\\
%sigma2
\dot{\sigma}_2 &=&   \Big(1 - 3{\cal N} \Big) \sigma _2
- {\textstyle \frac13 }  \, {\cal N}\left(   \big(2\nu_2 - \nu_1 - \nu_3 \big) \nu_2 -  \big(  \nu_1- \nu_3  \big)^2
 \right)
 ,\\
%nu_1
\label{nu1-ev-ul-app}
\dot{\nu}_1 &=& \Big( 1 + {\cal N} \big(   2   \sigma_1 - 1 \big) \Big) \nu_1  
,\\
%nu2
\dot{\nu}_2 &=&  \Big( 1 + {\cal N} \big(   2   \sigma_2 - 1 \big) \Big) \nu_2  
,\\
%nu3
\dot{\nu}_3 &=& \Big( 1 - {\cal N} \big( 2\sigma_1 + 2\sigma_2  + 1 \big)   \Big) \nu_3   
,\\
%W
\label{eq-w-ul}
\dot{\bar{W}}&=& -\Big( 3{\cal N} - 1 \Big) \bar{W}  -{\cal N} \,\bar{V}_{,\phi},
\end{eqnarray}
where $\sigma_i$ and $\nu_i$ $(i=1,2,3)$ denote the Hubble-normalized shear and spatial curvature tensors, respectively, and $\bar{W}$ is the Hubble-normalized scalar field velocity. 
Note that, by definition, the shear tensor $\bar{\Sigma}_{ab}$ is trace-free and hence $\sigma_3 = - \sigma_1 - \sigma_2$ is not an independent variable. 

The eigenvalue system~(\ref{eq-sigma1-ul}-\ref{eq-w-ul}) is subject to the Hamiltonian constraint, \begin{equation}
\label{constraintG-ul}
{\cal N}^{-1}  = 3 - \bar{V} - {\textstyle \frac13 } \Big(\nu_1^2+\nu_2^2+\nu_3^2\Big)
+ {\textstyle \frac16 } \Big( \nu_1+\nu_2+\nu_3 \Big)^2 ,
\end{equation}
and the CMC gauge condition that yields the lapse equation,
\begin{equation}
\label{Neqn-limit}
3{\cal N}^{-1} = 3 + \bar{W}^2   - \bar{V}(\phi) 
+  2\Big( \sigma_1^2 + \sigma_1\sigma_2 + \sigma_2^2  \Big).
\end{equation}

The fixed point solutions of the eigenvalue ODE  system have been identified in Ref.~\cite{Ijjas:2020dws} for the case of $\bar{V}_{,\phi}/\bar{V} \equiv {\rm \it constant}$. 
In the frozen coefficient approximation where we treat the system as one with constant coefficients for each fixed value of $\phi$, it is straightforward to verify, {\it e.g.}, by using a computer algebra system, that the ODE system admits the same type of fixed point solutions for $\bar{V}_{,\phi}/\bar{V} \neq {\rm \it constant}$ and changing adiabatically:
\begin{itemize}
\item if $\bar{V}>0$, there exist four spatially curved fixed point solutions (with at least one of the $\nu_i\neq0$); 
\item if $\bar{V}\simeq0$, there exists a spatially curved ({\it e.g.}, $\nu_1=\nu_2\neq0, \nu_3=0$) and anisotropic ($\sigma_1 = \sigma_2 = -1$) fixed point and a flat (all $\nu_i=0$) but anisotropic ($\sigma_1, \sigma_2\neq0$) Kasner-like (${\cal N}=1/3$) fixed point; 
\item if $\bar{V}<0 $, there exists a flat FRW (all $\sigma_i, \nu_i=0$) fixed point solution:
\begin{equation}
\label{fixed-point}
\bar{W} = - \frac{\bar{V}_{,\phi}}{\bar{V}},\quad 
\bar{V}= 3-\frac12 \left(\frac{\bar{V}_{,\phi}}{\bar{V}}\right)^2,\quad 
{\cal N}^{-1}=\frac12 \left(\frac{\bar{V}_{,\phi}}{\bar{V}}\right)^2.
\end{equation}
\end{itemize}

To show the stability of the flat FRW fixed point solution for the general case of $\bar{V}<0$ and $\bar{V}_{,\phi}/\bar{V} \neq {\rm \it constant}$, we linearize the eigenvalue system~(\ref{eq-sigma1-ul}-\ref{eq-w-ul}) around the fixed point~\eqref{fixed-point}: 
\begin{eqnarray}
%sigma1
\label{dsigma}
\delta \dot{\sigma}_i &=&   \left( 1 - 6\left(\frac{\bar{V}_{,\phi}}{\bar{V}}\right)^{-2} \right) \delta \sigma _i  
\\
%nu_1
\label{nu1-ev-ul-app-p}
\delta\dot{\nu}_i &=& \left( 1 - 2\left(\frac{\bar{V}_{,\phi}}{\bar{V}}\right)^{-2} \right) \delta \nu_i  
,\\
%delta-phi
\label{dphi}
\delta\dot{\phi} &=& - 2 \left(\frac{\bar{V}_{,\phi}}{\bar{V}}\right)^{-2} \delta \bar{W}  
,\\
%W
\label{dbarw}
\delta\dot{\bar{W}}&=& \left( 1 - 6\left(\frac{\bar{V}_{,\phi}}{\bar{V}}\right)^{-2} \right) \left( \delta \bar{W}   + \left(\frac{\bar{V}_{,\phi\phi}}{\bar{V}} - \left(\frac{\bar{V}_{,\phi}}{\bar{V}}\right)^2 \right) \delta \phi \right)
 ,
\end{eqnarray}
where we used Eqs.~(\ref{constraintG-ul}-\ref{Neqn-limit}), 
to eliminate the dependent variables  $\delta {\cal N} = - {\cal N}^2 \bar{W} \delta \bar{W}$ and $\delta {\bar V}=-  \bar{W} \delta \bar{W}$.
For the stability analysis, we consider sufficiently small environments around each field value $\phi_0$ such that the frozen coefficient approximation holds, {\it i.e.}, the coefficients of the linearized system~(\ref{dsigma}-\ref{dbarw}) can be viewed as constants.

The perturbations around the flat FRW fixed point decay as $t\to-\infty$, {\it i.e.}, the fixed point solution is a stable attractor, if:
\begin{equation}
\label{attractor-condition}
M_{\rm Pl} \left|\frac{\bar{V}_{,\phi}}{\bar{V}}\right| > \sqrt{6} \quad {\rm and} \quad 
\frac18 \left(\left(\frac{\bar{V}_{,\phi}}{\bar{V}}\right)^2 - 6M_{\rm Pl}^{-2} \right) \geq \frac{\bar{V}_{,\phi\phi}}{\bar{V}} - \left(\frac{\bar{V}_{,\phi}}{\bar{V}}\right)^2 
\geq 0
,
\end{equation} 
as the scalar field is rolling downhill.
The key difference from the case with $M_{\rm Pl}|\bar{V}_{,\phi}/\bar{V}| = {\rm \it constant}$ is the term $\propto \delta\phi$ in Eq.~\eqref{dbarw}. Due to the presence of this term, the equations for $\delta \phi$ and $\delta \bar{W}$ are coupled, which results in the second condition in Eq.~\eqref{attractor-condition} that bounds $\bar{V}_{,\phi\phi}/\bar{V}$ both from above and below.

As we have seen in the examples presented above in Sec.~\ref{examples-smoothing}, due to the extraordinary rapidity of smoothing, the potential can have an overall power-law shape if it exhibits exponential behavior for some time, sourcing a sufficiently large and nearly constant $M_{\rm Pl}|\bar{V}_{,\phi}/\bar{V}|$. 

As to how long this requirement must be met, depends on the initial conditions. For generic initial conditions that lie far outside the perturbative regime of flat FRW geometries, the whole of spacetime is typically smoothed within $N_{\Theta} \simeq$ 10-20 $e$-folds of contraction of the inverse mean curvature $\Theta$, if $M_{\rm Pl}|\bar{V}_{,\phi}/\bar{V}|\gtrsim5$. Once the flat FRW state is reached, the system stays there and spacetime continues to slowly contract for any negative potential. The overall potential shape only determines how much faster the Hubble radius shrinks than the scale factor, {\it i.e.} by how much physical distances contract during 60 or more $e$-folds of contraction in the Hubble radius. The shallower the potential the more physical distances shrink for the same amount of contraction in the Hubble radius.

Although rapid and robust smoothing does not require an overall exponential (or super-exponential) potential, various theoretical arguments as to what might constitute a consistent theory of quantum gravity point towards steep exponentials or super-exponentials; see {\it e.g.}, Ref.~\cite{Obied:2018sgi}. For example, the `range constraint' \cite{Ooguri:2006in} states that for low-energy effective theories  to remain valid, the scalar field should roll over a range of $\Delta \phi \sim {\cal O}(1)$ (in reduced Planck units). In the case of a negative exponential potential as given in Eq.~\eqref{exponential-potential} with $\bar{V}_{,\phi}/\bar{V} = -1/m$, we can approximate $\Delta \phi$ using the fixed point solution as given in Eq.~\eqref{fixed-point}: 
\begin{equation}
\Delta \phi \simeq 2 m N_{\Theta},
\label{delta-phi}
\end{equation}
where $N_{\Theta}=t_0-t$ measures the number of $e$-folds of contraction of the inverse mean curvature $\Theta$, see Eq.~\eqref{time-def}. That is, during $N_{\Theta} \sim {\cal O}(100)$ $e$-folds of contraction, $\Delta \phi \sim {\cal O}(1)$ if $m/M_{\rm Pl} \sim {\cal O}(10^{-2})$. For such a potential, physical distances shrink by a factor of 3 during 60 $e$-folds of contraction in the Hubble radius.

In the cases where we observed that the system approaches the Kasner-like fixed point, the potential wasn't steep enough for the system to reach the flat FRW state. The same occurs for positive potentials or no potentials. In all of these cases, $\bar{V}$ rapidly becomes negligible and the dynamical behavior corresponds to that of a free scalar and, as shown in Ref.~\cite{Ijjas:2020dws}, the system approaches the Kasner-like fixed point. Note there is a continuous set of possible Kasner-like states (see, {\it e.g.}, Row 2 of Table~1 in Ref.~\cite{Ijjas:2020dws}) such that the dynamics can evolve to an inhomogeneous chaotic mixmaster state.  Scalars with such potentials are ill-suited microphysical sources for smoothing slow contraction.

\section{Graceful exit, boundedness and the uphill slope}
\label{sec:uphill}

All potentials considered thus far have in common that they are unbounded from below, meaning that the phase of smoothing slow contraction could last arbitrarily long and the Ricci curvature, the Hubble radius and energy density could get arbitrary large.  
However, in scenarios that involve a classical (non-singular) bounce, slow contraction must conclude {\it well} before the Hubble radius shrinks to the size of the Planck length; see, {\it e.g.}, \cite{Ijjas:2019pyf,Ijjas:2021zwv}. 

The transition from slow contraction to the bounce stage is only successful, though, if it preserves the homogeneity, isotropy and flatness over exponentially many super-Hubble scales (`graceful exit'). 
In the remainder of this paper, we construct realistic models of slow contraction that include an exit stage and use the tools of numerical relativity to explore whether and how the smoothing power of slow contraction is affected by modifying the scalar field potential to include the exit stage.

\subsection{General prescriptions for model building}

We continue considering scalar fields with canonical kinetic energy that are minimally-coupled to Einstein gravity. Realistic potentials that generate an extended phase of smoothing slow contraction that is robust to a wide range of initial conditions and lead to an end of slow contraction share three characteristic features: 
\begin{itemize}
\item a downhill slope with $M_{\rm Pl}|V_{,\phi}/V|>5$ to source $N_{\Theta}\sim 60$ or more $e$-folds of slow contraction;
\item a potential minimum that leads to an exit from slow contraction; and
\item an uphill slope that ends with a slightly positive plateau or a local minimum the height of which, $V_{\rm DE}$, corresponds to today's dark energy density. 
\end{itemize}

First, to account for today's observable universe, slow contraction must last sufficiently long. In particular, to generate the primordial fluctuations observed in the CMB, spacetime must converge to the homogeneous, isotropic and flat FRW state {\it before} the last $N_{\Theta} \sim 60$ $e$-folds of slow contraction of the inverse mean curvature $\Theta$, see, {\it e.g.}, Ref.~\cite{Ijjas:2020dws}. 
For general initial conditions, earlier numerical studies \cite{Cook:2020oaj,Ijjas:2020dws,Ijjas:2021gkf,Ijjas:2021wml} as well as our results presented in the previous Section demonstrate that slow contraction is a robust and rapid smoother, typically reaching the flat FRW state within the first 10-20 $e$-folds of contraction in the inverse mean curvature, $\Theta$, even if the initial conditions lie well outside the perturbative regime of the flat FRW state. Overall then, the downhill slope of the potential must be sufficiently long to accommodate $N_{\Theta}\sim80$ or more $e$-foldings of slow contraction.

Second, to exit after $N_{\Theta}\sim80$ or more $e$-foldings of slow contraction, the kinetic energy density of the scalar field must increase relative to its potential energy density such that the potential energy density becomes negligible relative to the total energy density, {\it i.e.}, $\bar{V}\to0$, causing the equation of state $\epsilon$ to approach 3 from above and ending slow contraction.

During the smoothing phase, the kinetic and potential energy densities increase at the same rate, as can be seen from Eq.~\eqref{fixed-point}. In principle, the relative contribution of the potential energy density, $\bar{V}$, can become negligible in two distinct ways: with a negative potential that is bounded from below or with a negative potential that is unbounded from below if its slope becomes too shallow to continue sourcing slow contraction, {\it i.e.}, $M_{\rm Pl}|V_{,\phi}/V|\leq\sqrt{6}$. 

However, if smoothing slow contraction is sourced by a scalar potential with $M_{\rm Pl}|V_{,\phi}/V|\leq\sqrt{6}$ (after the smooth FRW state is reached) and which is unbounded from below, the scalar field must explore exponentially large field ranges. Arguments from fundamental physics \cite{Ooguri:2006in,Agrawal:2018own,Ooguri:2018wrx,Grimm:2018ohb,Bernardo:2021wnv} suggest that an effective theory can only be consistent with  quantum gravity if the scalar field explores a finite range of ${\cal O}(1)$ in reduced Planck units. 
For this reason, to prevent the scalar from rolling over field ranges $\gtrsim{\cal O}(1)$, we only consider potentials that are bounded from below.  

Given the requirement of $N_{\Theta}\sim80$ or more $e$-foldings of slow contraction, from Eq.~\eqref{delta-phi}, we can estimate the location of the minimum, $\phi_{\rm min}$, independently of the exact shape of the potential in that region:
\begin{align}
    \phi_{\rm min} \simeq \phi_0 - 2 \,N_{\Theta} \,m;
\label{eq:phi_e}
\end{align}
here we use the fact that $\epsilon$ rapidly decreases and approaches 3 from above once the minimum is passed.
After that point, the scalar field kinetic energy continues being blue-shifted by the large Hubble anti-friction such that the field starts rolling uphill instead of remaining at the potential minimum. As the field is climbing uphill, its potential energy density cannot keep up with the growing kinetic energy density. The relative contribution of the potential to the total energy density continuously decreases and $\bar{V}$ eventually becomes negligible causing $\epsilon \to 3$.

Third, we require that the potential does not asymptote zero from below but instead has a small plateau or a small local minimum with $V_{\rm DE}>0$, corresponding to the currently measured value of the dark energy density. In the case of a local minimum, the plateau is replaced by  a small positive barrier followed by the small positive local minimum. In this case, the strongly blue-shifting scalar field kinetic energy pushes the field over the barrier such that the scalar reaches the positive local minimum, starts to oscillate, decaying into the particles of the standard model and reheating the universe. Along the way, spacetime transitions to the hot expanding phase through a cosmological bounce that can take place before or after reheating; see {\it e.g.}, Ref.~\cite{Ijjas:2019pyf}.   

\subsection{Worked Examples}

Next, we construct sample potentials that lead to an exit from slow contraction after $N_{\Theta}\sim80$  $e$-folds to test the effect on the robustness of smoothing. 

We are interested in potentials of the form
\begin{align}
V(\phi) \simeq  
 \begin{cases}
-V_0 e^{-\phi/m} + V_{\rm DE}, &\text{if $\phi\gg \phi_{\rm min} $},
\\
-\tilde{V}_0 e^{ \phi/\tilde{m}}  + V_{\rm DE}, &\text{if $\phi\ll _{\rm min}$},
\end{cases} 
\label{eq:V_exit}
\end{align}
such that  $V$ has a local minimum at $\phi_{\rm min}$ with the downhill slope ($\phi\gg\phi_{\rm min}$) exponentially decreasing at the rate of $m^{-1}$ and the uphill slope ($\phi\ll\phi_{\rm min}$) exponentially increasing at the rate of $\tilde{m}^{-1}$, and the constant $V_{\rm DE}>0$ generating a small positive plateau. (In the case that the plateau is replaced by a barrier followed by a small positive local minimum, we found that there was no change to the robustness and rapidity of smoothing during slow contraction.)

A concrete example that we use in our numerical simulations is given by
\begin{equation}
 V(\phi) = -V_0 e^{-\phi/m} \times \frac12 \times \left( \tanh \left(\frac{\phi-\phi_{\ast}}{m_{\ast}}\right) + 1\right) + V_{\rm DE},
 \label{eq:V_tanh}
\end{equation}
where $m_{\ast}<2m$ such that $V(\phi)$ has a minimum 
at
\begin{equation}
\phi_{\rm min}=\phi_{\ast} + \frac{1}{2}\times\ln\left(2\frac{m}{m_{\ast}}-1\right)\times m_{\ast}.
\end{equation} 
The exponential uphill slope defined in Eq.~\eqref{eq:V_exit} is fixed through
the coefficient $\tilde{V}_0=V_0 e^{-2\phi_{\ast}/m_{\ast}}$ and exponent $\tilde{m}^{-1}=2(m/m_{\ast}-1/2)m^{-1}>0$ .

All simulations presented below have in common that, when defining the parameters for the scalar field and its potential, we chose the least favorable conditions for slow contraction to start. For example, we set 
\begin{align}
&V_0 = 0.1 \,M_{\rm Pl}^2 \Theta_{0}^{-2},&
&V_{\rm DE}/V_0 = 1,&
&\phi_0 = 0, &
\label{eq:ic_pot}
\end{align}
such that the field starts at $V(\phi)=0$, eliminating the possibility of smoothing in regions that might undergo a brief phase of accelerated expansion sourced by $V_{\rm DE}$ before transitioning to contraction, and eliminating the possibility that the initial conditions are biased towards smoothing slow contraction because the field starts out with a  negative value of the potential. 

In addition, we set
\begin{align}
&m_{\ast}=0.5\,m ,&
&\phi_{\ast} =  - 2 \,N_{\Theta} \,m   &
&{\rm and}&
&N_{\Theta}=80,   &
\label{eq:ic_pot2}
\end{align}
which makes the uphill slope much steeper ($\tilde{m}^{-1}=3m^{-1}$) than the downhill slope decreases ($m^{-1}$) and we allow at most $N_{\Theta}\simeq20$ $e$-foldings of contraction to reach complete smoothing. 
It is essential for the background to reach complete smoothing during the first 20 $e$-folds. Otherwise, it would not be flat FRW during the last 60 $e$-folds as required to generate the nearly scale invariant spectrum of temperature  fluctuations observed in the CMB.

Furthermore, as specified in the next two sections, we choose the initial shear and scalar field velocity contributions such that the resulting initial state is far outside the perturbative regime of flat FRW spacetimes.

\subsubsection{Inhomogeneities along one spatial axis}
\label{subsec:1D}

We start with considering initial conditions that involve inhomogeneities along a single spatial direction. As we will see below in Sec.~\ref{sec:(2+1)D}, this symmetry restriction leads to the least favorable initial conditions for slow contraction to start. Nevertheless, we found that the remarkable robustness and rapidity of smoothing remain unaffected by altering the shape of the scalar field potential to include a large negative minimum such that the scalar field exits slow contraction and, climbing uphill, the field ends up on a small positive plateau corresponding to today's dark energy density.

To facilitate comparison with the case of a pure negative exponential, we employed the same parameters to specify the initial shear and scalar field velocity contributions defined in Eqs.~(\ref{eq:z_ab0}) and (\ref{eq:q0}), respectively, {\it i.e.},
\begin{align}
\label{eq:ic_1d}
    a_{1,2} &= 0.5,
    &b_1 &= -0.15,
    &b_2 &= 1.8, 
    &\xi &= 0.01,
    &\kappa_{1,2} &= 0.01,
    &c_{1,2} &= 0,
    \\ 
    \label{eq:ic_1d-2}
    q_x &= 0.1\,m^{-1},
    &q_y &= 0,
    &d_x &= -0.15,
    &d_y &= 0,
    &m_x &= 1,
    &m_y &= 0.
\end{align}
We varied the parameters determining the average initial velocity $Q_0$ of the scalar and the steepness of the downhill slope of the potential $|V_{,\phi}/V| =m^{-1}$ because these are the only two quantities that can affect whether and how complete smoothing is reached.

%%%FIGURE-1D-PHASE-DIAGRAM
\begin{figure}[tb]
  \centering
\includegraphics[width=.575\textwidth]{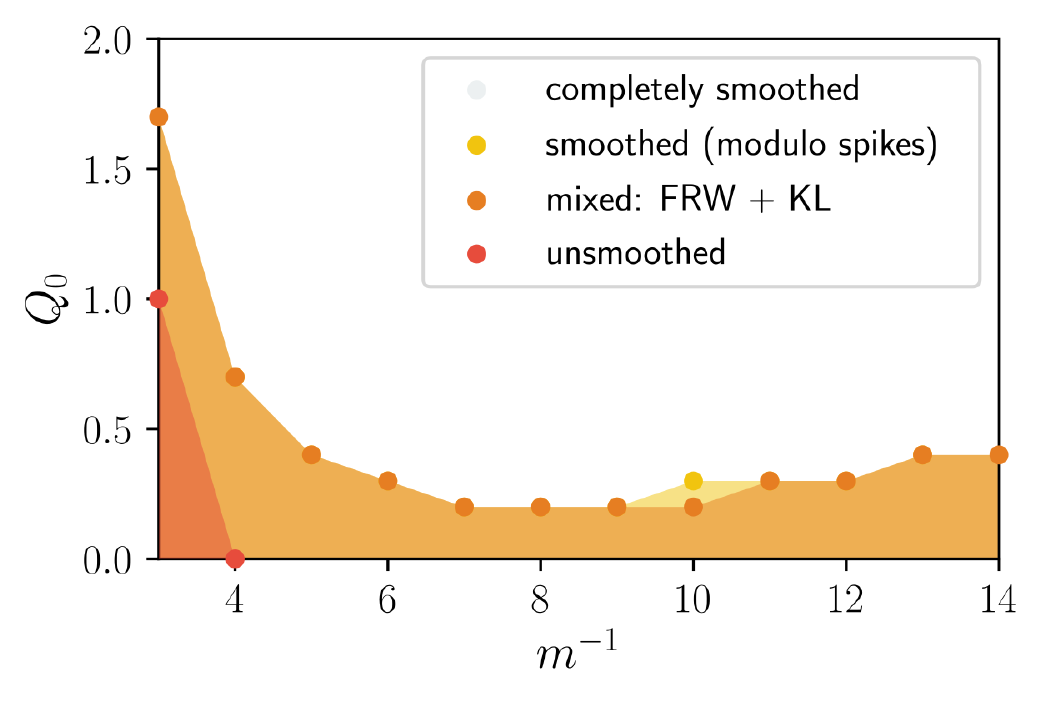}
  \caption{Phase diagram showing possible final states corresponding to the potential given by Eq.~\eqref{eq:V_tanh} with parameters defined in Eqs.~(\ref{eq:ic_pot}-\ref{eq:ic_pot2}) and initial conditions defined in Eqs.~(\ref{eq:ic_1d}-\ref{eq:ic_1d-2}). The diagram shows the final states as a function of the inverse mass scale $m^{-1}$ associated with the downhill slope of the scalar field  potential  and its mean initial velocity $Q_0$. As in the case of a pure exponential, the entire region relevant to cyclic and bouncing models ($Q_0>0$ and $m<0.25$)  converges completely or to an exponential degree (as measured by proper volume) to the flat FRW attractor solution.}
 \label{fig:phase-1d}
\end{figure}
Figure~\ref{fig:phase-1d} summarizes our findings in form of a phase diagram first introduced in Ref.~\cite{Ijjas:2020dws}. Each point of the diagram corresponds to a simulation with a different combination of $(m^{-1},Q_0)$. The large white region represents all simulations that lead to complete smoothing. In each of these simulations, the flat FRW state is reached everywhere within $N_{\Theta}=20$ $e$-folds of contraction in the inverse mean curvature $\Theta$; the field passes through the potential minimum by $N_{\Theta}\simeq80$ and reaches the positive potential plateau by $N_{\Theta}\simeq90$. 

As seen in earlier studies involving pure negative exponentials, the smoothing power decreases with smaller mean initial velocities and the system eventually starts to exhibit spiking behavior. For even smaller initial velocities ($Q_0\leq0$), spacetime is not smoothed completely but ends up with a mixture of flat FRW and Kasner-like states. As pointed out in Refs.~\cite{Cook:2020oaj,Ijjas:2020dws}, the reason is that, if the mean initial velocity is uphill directed ($Q_0<0$), there will always be some regions in which the scalar's potential energy cannot catch up to its kinetic energy such that, in those regions, the scalar field behaves like a free field, resulting in convergence to an ultralocal, anisotropic Kasner-like state. 

Only a small region of parameter space leads to final states that are not completely smooth: a final state that is nowhere smooth can only be obtained if the characteristic mass scale $m$ of the scalar field potential is large, i.e., $m\geq0.25M_{\rm Pl}$, corresponding to a too shallow downward slope. 
Note, in addition, that $Q_0$ is measured in units of the mean curvature $\Theta_0^{-1}$ at the initial time $t_0$ which is vanishingly small in Planck units. For the purposes of illustration, we have intentionally limited the phase diagram to the tiny region where $Q_0\sim {\cal O}(1)\,\Theta_0^{-1}$. The overwhelming majority of parameter space with $Q_0\gg\Theta_0^{-1}$ consists entirely of completely smooth final states.

Comparing to the case of a pure exponential potential with similar initial conditions as, {\it e.g.}, in Phase Diagram V of Ref.~\cite{Ijjas:2020dws}, it becomes clear that the robustness of smoothing remains unaffected by altering the shape of the scalar field potential to include an exit from the smoothing phase.  Even though we chose different initial conditions, the phase boundary of the mixed state is only shifted slightly in the diagram. As emphasized above, these differences  are nominal since the parameter space in the phase diagrams covers a vanishingly small part of the entire ($Q_0\geq0$) parameter space. In both cases, complete smoothing is reached except for a tiny region where $Q_0< {\cal O}(1)\,\Theta_0^{-1}$.

As a matter of fact, the results found when bounding the potential from below are a further manifestation of the robustness and rapidity of smoothing through slow contraction. First, the robustness to initial conditions remains unaffected even though the system has less time -- at most $N_{\Theta}\simeq 80$ $e$-folds of smoothing phase before reaching the potential minimum -- compared to earlier simulations with unbounded potentials that were run for $N_{\Theta}\simeq200$ or more $e$-folds.  

%
%%FIGURE-epsilon_phi
\begin{figure*}[t!]
  \centering
\includegraphics[width=.875\textwidth]{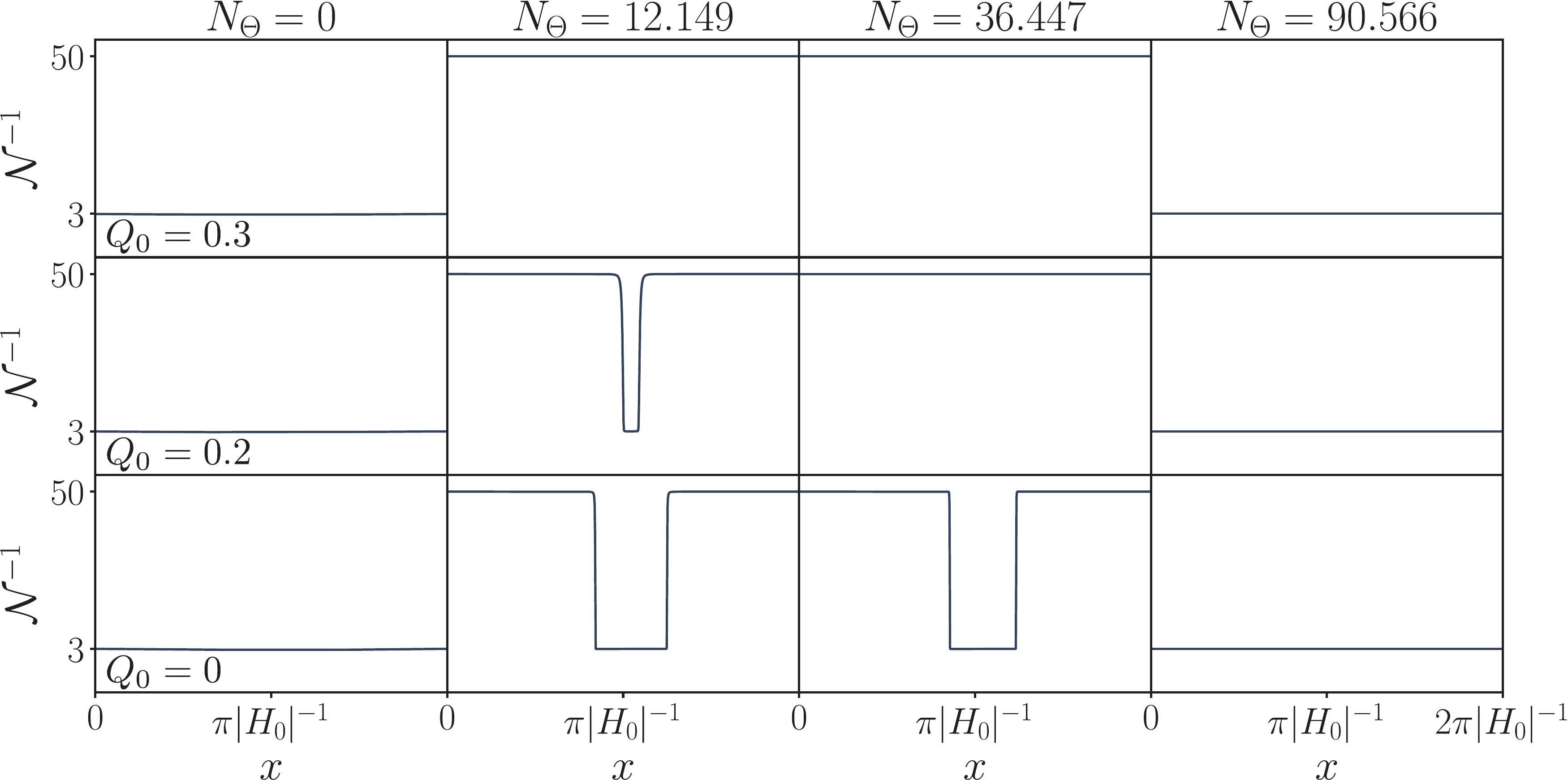}
  \caption{Snapshots of the evolution of the inverse Hubble-normalized lapse ${\cal N}^{-1}$ corresponding to the potential in Eq.~\eqref{eq:V_tanh} with $m=0.1M_{\rm Pl}$ and different values of the mean initial velocity $Q_0$. Each row represents a different final state: complete smoothing by $N_{\Theta}\simeq12$  (upper row), complete smoothing modulo spikes by $N_{\Theta}\simeq 36$ (middle row) and a mixed final state (lower row). Once the scalar field reaches the minimum of the potential at $N_{\Theta}\simeq 80$, the field starts climbing uphill due to the large negative Hubble anti-friction. The smoothing phase ends by $N_{\Theta}\simeq90$ when the field reaches the positive potential plateau.}
\label{fig:calN-potential-w-end}
\end{figure*}
Second and more importantly, once the flat FRW state is reached, deviations from flatness, isotropy and homogeneity do not regrow after the system exits the smoothing stage.  
The reason why deviations do not regrow can be understood by linearizing the evolution equations listed in Appendix~\ref{app:scheme} around the flat FRW state. The perturbed geometric variables  $\delta\bar{E}_a{}^i, \delta\bar{n}_{ab}, \delta\bar{A}_b, \delta\bar{S}_a$ all share the same scaling behavior, {\it i.e.},
 \begin{equation}
 \delta\bar{E}_a{}^i, \delta\bar{n}_{ab}, \delta\bar{A}_b, \delta\bar{S}_a \propto e^{(1-{\cal N})t},
 \end{equation}
and the components of the linearized shear tensor scale as 
 \begin{equation}
 \delta\bar{\Sigma}_{ab}\propto e^{(1-3{\cal N})t}.
 \end{equation}
As slow contraction proceeds, $t<0$ decreases and all perturbations get suppressed exponentially because ${\cal N}$ remains below $1/3$ even after the scalar field reaches the potential minimum and approaches the positive potential plateau, as illustrated in Fig.~\ref{fig:calN-potential-w-end}. As the field climbs uphill,  ${\cal N}\to 1/3$ from below. Once the field arrives on the potential plateau, ${\cal N} = 1/3$, meaning that the shear perturbations remain exponentially suppressed, they neither shrink nor grow, while the other perturbations continue to shrink exponentially.

Third, even in the case of simulations where the final state is a mix of flat FRW and Kasner-like regions, the proper 3-volume of Kasner-like regions is exponentially suppressed relative to the 3-volume occupied by regions with flat FRW geometry. As shown in Refs.~\cite{Garfinkle:2008ei,Ijjas:2020dws}, the proper 3-volume of a region corresponding to a segment with length $\Delta x$ is $\propto |\tau|^{3{\cal N}}$, where $\tau$ is the proper FRW time given through ${\rm d}\ln a(\tau)/{\rm d}\tau\equiv(1/3)e^{-t}$. Note that $\tau\to0$ as $t\to-\infty$ (see, Eq.~\ref{time-def}).
%REFER TO EQ.2.3
As shown in Fig.~\ref{fig:calN-potential-w-end}, during the smoothing phase ($10 \lesssim N_{\Theta} \lesssim80$), ${\cal N} = 2m^2 \ll1$ in the segments that contract slowly while ${\cal N}=1/3$ in Kasner-like regions. That is, as $\tau\to0$, the proper volume of Kasner-like regions shrinks at an exponentially faster rate than the proper volume of the flat FRW regions. After the field reaches the bottom of the potential, all regions start contracting at the same rate $\propto |\tau|$. As a result, the volume ratio remains the same with the Kasner-like regions occupying exponentially less proper volume.

The conclusions presented here were drawn from one specific sample potential. Performing simulations with different choices of $m/m_{\ast}$, ranging from  steep ($m/m_{\ast}=2$) to very shallow ($m/m_{\ast}=0.6$), we confirmed that the results presented above are insensitive to the choice of $m/m_{\ast}$. 
In particular, simulations with $m/m_{\ast}=0.6$ lead to the same phase diagram as the one depicted in Fig.~\ref{fig:phase-1d}.

\subsubsection{Inhomogeneities along two spatial axes}
\label{sec:(2+1)D}

%Figure 2D-omegas
\begin{figure*}[t!]
  \centering
\includegraphics[width=.9\textwidth]{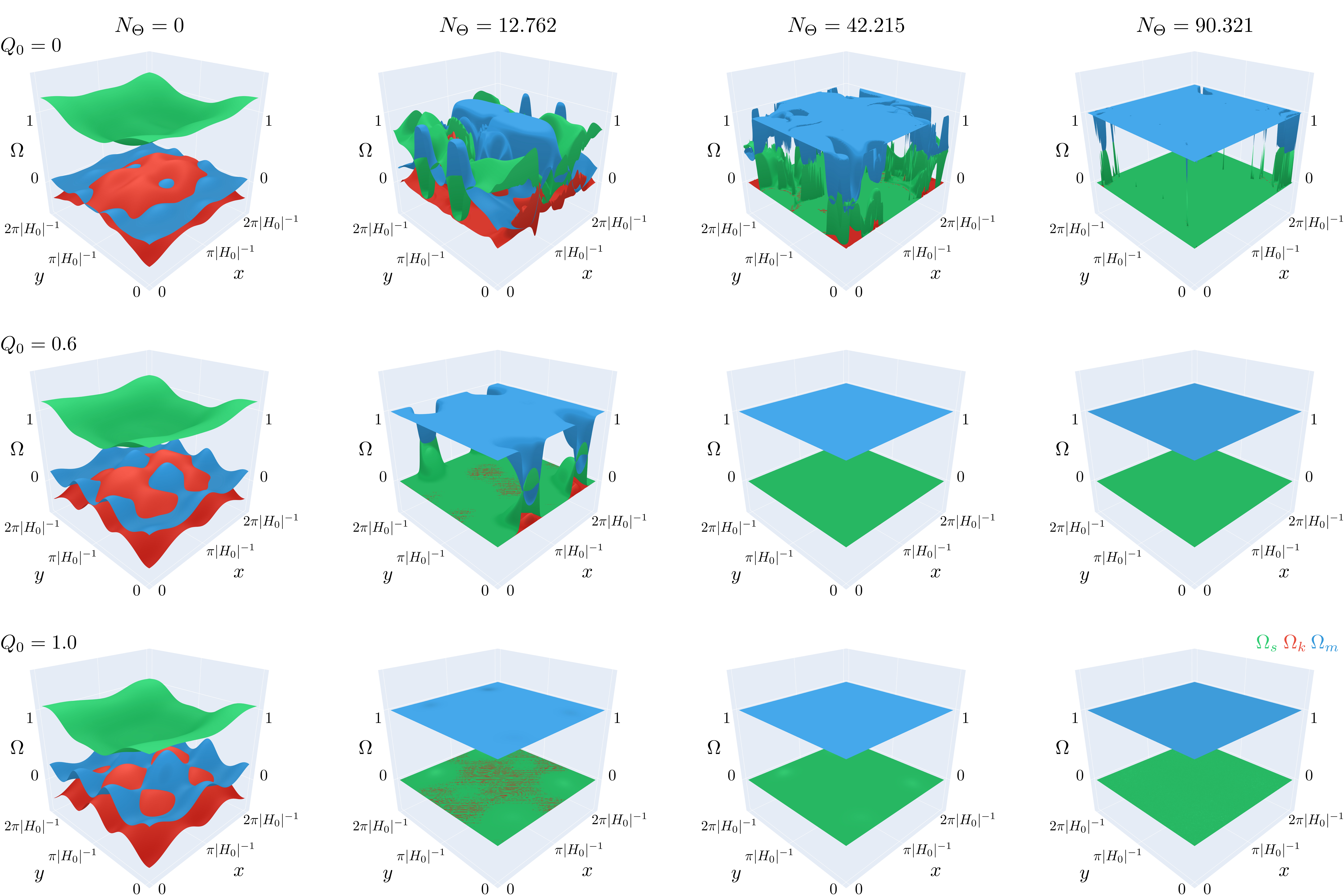}
  \caption{Snapshots of the Hubble-normalized energy density in shear $\Omega_s$ (green), curvature $\Omega_k$ (red), and scalar field $\Omega_{\phi}$ (blue) as a function of $N_{\Theta}$ corresponding to the potential defined in Eq.~\eqref{eq:V_tanh} with $m=0.25$ for three different values of the mean initial velocity of the scalar field.  These result in different smoothing outcomes: a mixed state of flat FRW and Kasner-like regions if $Q_0=0$ (upper row) , complete smoothing modulo spikes if $Q_0=0.6$ (middle row) and complete smoothing if $Q_0=1.0$ (lower row). 
 The initial state is highly curved, inhomogeneous and anisotropic along two spatial dimensions, as specified in Eq.~\eqref{eq:ic_2d}. }
\label{fig:omega-2d-06}
\end{figure*}
As a last step, we investigate the effect of inhomogeneities and anisotropies along two spatial axes.  The parameters fixing the initial shear and scalar field velocity contributions defined in Eqs.~(\ref{eq:z_ab0}) and (\ref{eq:q0}), respectively, are specified as follows:
\begin{align}
\label{eq:ic_2d}
a_{1,2} &= 0.5,
&b_1 &= -0.15,
&b_2 &= 1.8, 
&c_{1,2} &= 0.5,
&\kappa_{1,2} &= 0.01,
&\xi &= 0.01,
\\ \label{eq:ic_2d-2}
q_x &= 0.1m^{-1},
&q_y &= 0.1m^{-1},
&d_x &= -0.3,
&d_y &= -0.45,
&m_x &= 2,
&m_y &= 3.
\end{align}
To facilitate comparison with the 1D case presented in the previous Section~\ref{subsec:1D}, we kept the parameters determining the shape of the scalar field potential given in Eqs.~(\ref{eq:ic_pot}-\ref{eq:ic_pot2})
as well as all 1D quantities, $a_{1,2}, b_{1,2}$, $\kappa, \xi$ and $q_x$, given in Eqs.~(\ref{eq:ic_1d}-\ref{eq:ic_1d-2}) the same with the exception of the mode number of the initial scalar field velocity, $m_x$, that we increased from $1$ to $2$ to allow for higher mode spatial variations in the $x$-direction. In addition, to relax the symmetry restriction forbidding inhomogeneities and anisotropies along a second spatial direction, we assigned the parameters $c_{1,2}, q_y$ and $m_y$ non-zero values.

Figure~\ref{fig:omega-2d-06} illustrates three different evolutions corresponding to a potential with downhill slope $m^{-1}=4$ for three different values of the mean initial velocity $Q_0$: The final state is flat FRW everywhere if $Q_0=1$; flat FRW modulo spiking if $Q_0=0.6$ and a mix of flat FRW and Kasner-like regions if $Q_0=0$.
The difference to the 1D case is striking: As can be seen from Fig.~\ref{fig:phase-1d},  for $m=0.25$, $0\leq Q_0\leq0.7$ leads to a mixed state of flat FRW and Kasner-like regions while only $Q_0>0.7$ leads to complete smoothing, in the 2D case, all $0\leq Q_0<0.6$ lead to a mixed state and all $Q_0\geq0.6$ lead to complete smoothing (with or without spiking). This finding suggests that releasing the symmetry restriction on the initial conditions enhances the robustness of slow contraction, yielding further evidence for the conjecture in Ref.~\cite{Ijjas:2021wml} that the Kasner-like fixed point has a small basin of attraction.

\begin{figure}[tb]
  \centering
\includegraphics[width=.55\textwidth]{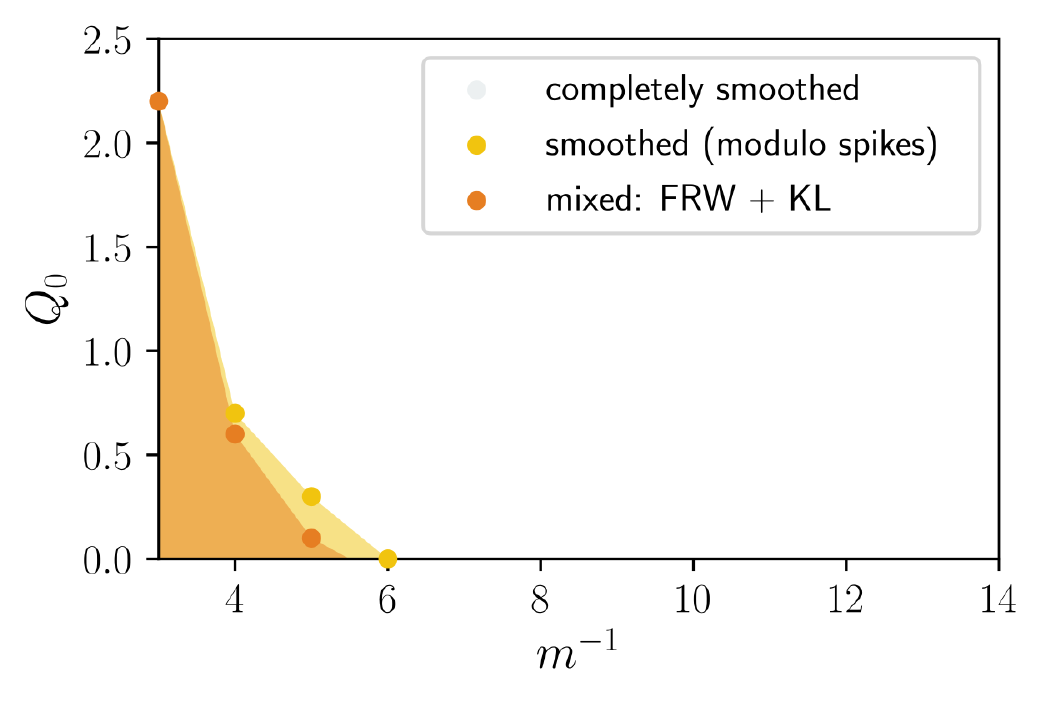}
  \caption{Phase diagram showing possible final states corresponding to the potential given by Eq.~\eqref{eq:V_tanh} with parameters defined in Eqs.~(\ref{eq:ic_pot}-\ref{eq:ic_pot2}) and initial conditions defined in Eqs.~(\ref{eq:ic_2d}-\ref{eq:ic_2d-2}) with inhomogeneities and anisotropies along two spatial directions. The diagram shows the final states as a function of the mean initial scalar field velocity $Q_0$ and the inverse mass scale $m^{-1}$ corresponding the downhill slope of the scalar field  potential. The entire region relevant to cyclic and bouncing models ($Q_0>0$)  converges completely or to an exponential degree (as measured by proper volume) to the flat FRW attractor solution. 
}
 \label{fig:phase-2d}
\end{figure}
The phase diagram in Figure \ref{fig:phase-2d} gives further supports the conjecture: In addition to the absence of the final state that is nowhere smooth, a comparison with Figure \ref{fig:phase-1d} shows that all phase boundaries are located below the ones from the one-dimensional simulations.

\section{Conclusion}
\label{sec:concl}

In this paper, we presented the first systematic study of more realistic 
scalar field potentials that can be incorporated in bouncing cosmologies. 
In particular, whereas previous studies only considered the slow contraction phase induced by purely negative exponential potentials that are unbounded below, here we have explored a much wider range of potentials that can also accommodate the graceful exit, bounce, and reheating stages that initiate the subsequent hot expansion phase. 

Using the methods of numerical relativity, we confirmed that a  large family of negative power-law, exponential and super-exponential potentials can source a smoothing slow contraction phase that rapidly converges to the homogeneous, isotropic and flat FRW state for a wide range of initial conditions, including those that lie far outside the perturbative regime of flat FRW geometries. Even though classically, once the flat FRW fixed point is reached, the state remains smooth whether the potential is power-law, exponential or super-exponential, arguments based on quantum gravity considerations suggest that potentials with an exponential or super-exponential downhill slopes are favorable because we find that, in those cases, the scalar field traverses a range of ${\cal O}(1)M_{\rm Pl}$ or less. 

In addition, we found that bounding the scalar potential from below to end the smoothing phase does not affect the robustness and rapidity of smoothing. Having reached the negative potential minimum, the scalar field continues to climb uphill until it reaches a small positive plateau or settles in a small positive local minimum to reheat the universe.

Most importantly, we established ultralocality sets in rapidly for a remarkably wide range of potentials, even in cases where the slow contraction phase lasts for a limited time.   The fact that gradient terms quickly become negligible is the main reason why altering the scalar potential to include an end to slow contraction has no affect on the robustness and rapidity of smoothing: The ultralocal state is typically reached within ${\cal O}(1)$ $e$-folds of contraction in the inverse mean curvature $\Theta$. Then, since gradients no longer affect the evolution, each spacetime point individually approaches the flat FRW attractor fixed point as long as the potential is sufficiently steep ({\it i.e.}, $M_{\rm Pl}|V_{,\phi}/V|\gtrsim 5$).   Once ultralocality is reached, spacetime remains flat FRW independent of the existence of a potential minimum and subsequent uphill slope leading to a small positive plateau or a small positive local minimum separated by a barrier.

Our numerical scheme is adapted to spacetimes where the mean curvature evolves monotonically. In particular, we did not consider cosmologies where slow contraction is preceded by a phase of dark energy expansion, as is the case in cyclic cosmologies. Extending our results to these scenarios is the focus of ongoing work.

\subsection*{Acknowledgements}
We thank Paul J. Steinhardt and Frans Pretorius for helpful comments and discussions. 
The work of A.I. is supported by the Simons Foundation grant number 947319. T.K. thanks Princeton University for hospitality, where parts of this work were completed.

\appendix

\section{Numerical scheme: evolution and constraint equations}
\label{app:scheme}

Using Hubble-normalized variables defined in Eqs.~(\ref{cal-N-def}-\ref{Vbar-def}) and imposing the frame and coordinate gauge conditions described in Sec.~\ref{sec:scheme}, the Einstein-scalar field equations (\ref{E-eq1}-\ref{E-eq2}) in orthonormal tetrad form  yield a hyperbolic system of evolution equations:
\begin{align}
%E_a^i
\label{eq-E-ai-Hn}
&\partial_t \bar{E}_a{}^i = - \Big({\cal N} - 1 \Big) \bar{E}_a{}^i - {\cal N} \,\bar{\Sigma}_a{}^b \bar{E}_b{}^i 
,\\
%Sigma_ab
\label{eq-sigma-ab}
&\partial _t \bar{\Sigma}_{ab} = - \Big( 3 {\cal N} - 1 \Big) \bar{\Sigma}_{ab}
- {\cal N} \Big( 2 \bar{n}_{\langle a}{}^c\, \bar{n}_{b \rangle c}
- \bar{n}^c{}_c \bar{n}_{\langle ab \rangle} 
- \bar{S}_{\langle a} \bar{S}_{b \rangle} \Big)
+ \bar{E}_{\langle a}{}^i\partial _i \Big(\bar{E}_{b \rangle}{}^i\partial _i {\cal N}\Big) 
\\
&\qquad\; - {\cal N} \left( \bar{E}_{\langle a}{}^i \partial_i \bar{A}_{b \rangle}
-  \epsilon^{cd}{}_{(a} \Big( \bar{E}_c{}^i\partial_i \bar{n}_{b)d} - 2 \bar{A}_c \bar{n}_{b )d} \Big) 
  \right)
+  \epsilon^{c d}{}_{(a} \bar{n}_{b ) d} \bar{E}_c{}^i \partial_i{\cal N}
+ \bar{A}_{\langle a} \bar{E}_{b \rangle}{}^i \partial_i {\cal N} 
\nonumber
,\\
%n_ab
\label{eq-n-ab}
&\partial _t \bar{n}_{ab} = - \Big({\cal N} - 1 \Big) \bar{n}_{ab} 
+ {\cal N} \Big( 2  \bar{n}_{(a}{}^c \bar{\Sigma}_{b)c}
-\epsilon^{cd}{}_{( a} \bar{E}_c{}^i \partial _i \bar{\Sigma} _{b ) d} \Big)   
- \epsilon^{cd}{}_{( a} \bar{\Sigma}_{b) d} \bar{E}_c{}^i \partial _i {\cal N} 
,\\
%A_a
\label{eq-A-a}
&\partial _t \bar{A}_a = - \Big( {\cal N} - 1 \Big)\bar{A}_a 
- {\cal N} \Big( \bar{\Sigma} _a{}^b \bar{A}_b - {\textstyle \frac12} \bar{E}_b{}^i\partial _i \bar{\Sigma} _a{}^b \Big)  
- \bar{E}_a{}^i \partial _i{\cal N} 
+ {\textstyle \frac12} \bar{\Sigma} _a{}^b \bar{E}_b{}^i \partial _i {\cal N} 
,\\
%phi
\label{eq-phi-Hn}
&\partial_t \phi = {\cal N} \,\bar{W}
,\\
%W
\label{eq-w-Hn}
&\partial_t \bar{W} = - \Big(  3 {\cal N} -1 \Big) \bar{W} 
- {\cal N} \Big(\bar{V}_{,\phi}  + 2 \bar{A}^a \bar{S}_a -  \bar{E}_a{}^i \partial_i \bar{S}^a  \Big)
+ \bar{S}^a \bar{E}_a{}^i\partial _i {\cal N}
,\\
%S_a
\label{eq-barS-Hn}
&\partial_t \bar{S}_a = - \Big(  {\cal N} - 1 \Big) \bar{S}_a 
- {\cal N}\Big( \bar{\Sigma}_a{}^b \bar{S}_b - \bar{E}_a{}^i\partial _i \bar{W} \Big)
+  \bar{W} \bar{E}_a{}^i \partial_i {\cal N}, 
\end{align}
where angle brackets denote traceless symmetrization defined as $X_{\langle ab \rangle} \equiv X_{(ab)} - {\textstyle \frac13}X_c{}^c\delta_{ab}$.

The hyperbolic PDE system~(\ref{eq-E-ai-Hn}-\ref{eq-barS-Hn}) is coupled to the elliptic lapse equation 
\begin{eqnarray}
\label{Neqn}
&-& \bar{E}^a{}_i \partial^i \left(\bar{E}_a{}^j \partial _j {\cal N}\right) + 2 
\bar{A}^a \bar{E}_a{}^i\partial _i {\cal N} + {\cal N} \left(3 + \bar{\Sigma} _{a b}\bar{\Sigma}^{a b} + \bar{W}^2   - \bar{V} \right) = 
3,
\end{eqnarray}
which results from our choice of CMC slicing. The system of Eqs.~(\ref{eq-E-ai-Hn}-\ref{Neqn}) serves as our numerical scheme.

Furthermore, Eqs.~(\ref{eq-E-ai-Hn}-\ref{eq-barS-Hn}) propagate the following constraints:
\begin{align}
\label{constraintG}
& {\textstyle \frac12 } \bar{\Sigma}^{ab} \bar{\Sigma}_{ab}
+ {\textstyle \frac12 } \bar{n}^{ab} \bar{n}_{ab}
- {\textstyle \frac14 } ( \bar{n}^c{}_c)^2
 - 2 \bar{E}_a{}^i \partial _i \bar{A}^a 
+ 3 \bar{A}^a \bar{A}_a 
%\\+& 
+ {\textstyle \frac12 } \bar{W}^2 +  {\textstyle \frac12 } \bar{S}^a \bar{S}_a + {\bar V} = 3
%\nonumber
,\\
\label{constraintC}
& \bar{E}_b{}^i \partial _i {\bar \Sigma} _a{}^b
 - 3 {\bar \Sigma} _a{}^b \bar{A}_b - \epsilon _a{}^{b c} \bar{n}_b{}^d \bar{\Sigma}_{cd} - {\bar W} {\bar S}_a = 0,
,\\
\label{constraintJ}
& \bar{E}_b{}^i \partial _i \bar{n}^b{}_a + \epsilon^{bc}{}_a \bar{E}_b{}^i\partial _i \bar{A}_c - 2 \bar{A}_b \bar{n}^b{}_a =0,\\
\label{constraintS-phi}
& {\bar S}_a - {\bar E}_a{}^i\partial _i\phi = 0,\\
\label{constraintCOM}
& \epsilon^{bc}{}_a
\Big( \bar{E}_b{}^j \partial_j \bar{E}_c{}^i - \bar{A}_b \bar{E}_c{}^i \Big) - \bar{n}_a{}^d \bar{E}_d{}^i = 0.
\end{align}
As detailed above in Sec.~\ref{sec:init_cond}, we use the constraints to specify the initial conditions as well as for code testing, in particular, to ensure numerical convergence. 

\section{Einstein-scalar system in the ultralocal limit}
\label{app:ODE}

In the ultralocal limit ($\bar{E}_a{}^i, \bar{A}_a, \bar{S}_a \to 0$), the evolution equations~(\ref{eq-E-ai-Hn}, \ref{eq-A-a} and \ref{eq-barS-Hn}) and the constraints~(\ref{constraintCOM}) are trivially satisfied such that the coupled Einstein-scalar non-linear PDE system~(\ref{eq-E-ai-Hn}-\ref{eq-barS-Hn}) reduces to the following coupled linear ODE system:
\begin{eqnarray}
%Sigma_ab
\label{eq-sigma-ab-ul}
\dot{\bar{\Sigma}}_{ab} &=& -\Big( 3{\cal N} - 1 \Big) \bar{\Sigma}_{ab} - {\cal N} \Big( 
2 \bar{n}^c{}_{\langle a} \bar{n}_{b\rangle c } - \bar{n}^c{}_c \bar{n}_{\langle ab \rangle} \Big),\\
%n_ab
\label{eq-n-ab-ul}
\dot{\bar{n}}_{ab} &=&  -\Big( {\cal N} - 1 \Big) \bar{n}_{ab} + 2\, {\cal N} \,  \bar{n}^c{}_{(a} \bar{\Sigma}_{b )c} ,
,\\
%phi
\label{eq-phi-Hn-ul}
\dot{\phi} &=& {\cal N} \,\bar{W}
,\\
%W
\label{eq-w-Hn-ul}
\dot{\bar{W}} &=& - \Big(  3 {\cal N} -1 \Big) \bar{W} 
- {\cal N} \bar{V}_{,\phi},
\end{eqnarray}
 where dot denotes differenciation w.r.t. the time coordinate $t$ as defined in Eq.~\eqref{time-def}.

\bibliographystyle{plain}
\bibliography{bib_paper1}

\end{document}